\documentclass{article}
\usepackage{fullname}
\title{Extended\\Comment on Language Trees and Zipping}
\author{Joshua Goodman\\Microsoft Research\\One Microsoft Way\\Redmond, WA 98052}

\newcommand{\mi}[1]{\mbox{\it #1}}

\begin{document}

\maketitle

\begin{abstract}This is the extended version of a Comment submitted to
Physical Review Letters.  I first point out the inappropriateness of
publishing a Letter unrelated to physics.  Next, I give experimental
results showing that the technique used in the Letter is 3 times worse
and 17 times slower than a simple baseline. And finally, I review the
literature, showing that the ideas of the Letter are not novel.  I
conclude by suggesting that Physical Review Letters should not publish
Letters unrelated to physics.
\end{abstract}

A recent Letter to Physical Review Letters, ``Language Trees and
Zipping,'' by \namecite{Benedetto:02a} (available at {\tt
http://arXiv.org/\linebreak[0]abs/\linebreak[0]cond-mat/\linebreak[0]0108530})
is flawed in several ways.  First of all, the Letter had nothing to do
with physics, and instead belonged in a computer science journal, if
it deserved to be published at all.  Second of all, the actual results
are unimpressive: as I will show, the techniques used lead to 3 times
as many errors and are 17 times slower than a very simple baseline
model applied to a standard, similar problem.  Finally, the ideas in
the Letter are not even novel: they are well known to those in several
areas of computer science.

The actual paper is clearly unrelated to physics, and much more
closely related to areas of computer science such as Computational
Linguistics and Machine Learning, as can be seen simply by reading the
abstract of their paper, which I include here:
\begin{quote} 
In this Letter we present a very general method for extracting
information from a generic string of characters, e.g., a text, a DNA
sequence, or a time series. Based on data-compression techniques, its
key point is the computation of a suitable measure of the remoteness
of two bodies of knowledge. We present the implementation of the
method to linguistic motivated problems, featuring highly accurate
results for language recognition, authorship attribution, and language
classification.
\end{quote}
The actual technique used applied standard concepts from machine
learning, including the Minimum Description Length principle, and
actually used a standard compression algorithm (LZW), found in
programs like gzip.  Thus, the paper is neither an application to
physics, nor an application of physics.

Physics journals should not be publishing articles that have nothing
to do with physics.  Of course, it is completely reasonable to publish
applications of physics to other fields (both because this alerts
other physicists to the possibility of applying their knowledge and
because those in the field of interest may have difficulty
understanding the terminology or techniques).  It is also completely
reasonable to publish the use of non-physics techniques applied to
physics in a physics journal.  But this paper applies computer science
techniques (gzip!) to computer science problems.  This seems extremely
inappropriate. One might argue that the paper discusses entropy, a
concept taken from physics.  But the concept was taken from physics 50
years ago, at the dawn of computer science, and there is nothing
physics-specific in the use of entropy in this paper; indeed, the use
is entirely in the information theory/language modeling/computer
science meaning of the word.

Given this argument, I don't think it really matters whether or not
the paper is a good paper -- the point is, quite simply, that a good
paper that has nothing to do with physics does not belong in a physics
journal.  As it happens, the paper is a bad paper.  First of
all, there are many many well known techniques in Computer Science for
solving problems of this sort, some complex and some very simple.  I
decided to try the simplest, easiest to implement, most
straightforward algorithm I could think of, an algorithm known as
Naive Bayes.  I implemented both the zipping algorithm and a Naive
Bayes algorithm, and applied them to a similar, but standard problem,
20 Newsgroups, the goal of which is to determine the newsgroup from
which a posting came.  The zipping procedure is not even that much
simpler: it takes about 70 lines of perl script to write the code to
split the data into test and training, and an additional 50 to use the
zipping method, versus an additional 70 to implement Naive Bayes (a
difference of 20 lines is trivial.)  The zipping method was 17 times
slower and had 3 times as many errors as this simplest standard
computer science algorithm.  See Appendix A for details.

Furthermore, the ideas of this paper are very well known in areas of
computer science such as machine learning and statistical natural
language processing.  I'll cite just a few of the hundreds of papers
related to this area.  To give an idea how basic this paper is
compared to the state of the art, \namecite{Peskin:93a} showed that
similar ideas (Bayesian analysis, which is essentially equivalent to
MDL) could be applied to doing topic and speaker identification not on
text (which is easy) but on actual speech, a much harder problem.
\namecite{Lowe:94a} showed that they could use these techniques to do
language identification from a few seconds of speech.  These ideas are
now very well known: see for instance, all of Chapter 16 of
\namecite{Manning:99a} (a standard introductory textbook) -- a
chapter devoted to techniques for text classification, the problem
area discussed in this paper.  The chapter, of course, mentions Naive
Bayes classifiers, which are a simple application of the MDL principle
-- much simpler in fact than the algorithms in compression programs.
Given the simplicity and appropriateness of Naive Bayes, it's no
wonder that computer scientists use it instead of much more complex
compression programs.  See also page 515 where Manning and Schutze
briefly discuss Cheeseman {\em et al.'s} \shortcite{Cheeseman:88a}
work on using MDL for clustering -- essentially what was done in this
paper.  The only idea in this paper that is not very well known in the
field is the idea of actually using a standard compression tool to do
the classification.  Still, even this idea dates back (at least) to
1995, when Ken Lang and Rich Caruana tried out the idea for doing
newsgroup classification.  In the end though, they ended up using
Naive Bayes classifiers \cite{Lang:95a}.  They didn't bother to
publish the compression idea because they thought it was better viewed
as an interesting thought than as a serious classification method.
Still, the idea of using compress got around a bit: see an
introductory tutorial by a well known practitioner in this area, Tom
\namecite[page 11]{Mitchell:97a}.  Admittedly, however, this technique
is not that widely known, because computer scientists don't typically
try anything that crude -- in a couple of hours (or less), we can
build from scratch tools that work better than this.

Of course, this explains another reason why physics journals should
not be publishing computer science papers: they don't know the field,
or the reviewers, and so cannot distinguish the good from the bad,
this paper being a case in point.

Why am I so bothered by this paper?  Well, a big part of it has to do
with the amount of press coverage it has received.  The authors sent
out a press release that got published in Nature Science Update (see
{\tt http://www.nature.com/\linebreak[0]nsu/\linebreak[0]020121/\linebreak[0]020121-2.html}) and Wired
Magazine (see {\tt
http://www.wired.com/\linebreak[0]news/\linebreak[0]technology/\linebreak[0]0,1282,50192,00.html}) and picked
up by people such as the ACM (Association for Computing Machinery)
(see {\tt
http://www.acm.org/\linebreak[0]technews/\linebreak[0]articles/\linebreak[0]2002-4/\linebreak[0]0208f.html\#item14}), who
perhaps should have known better than to trust stuff from a physics
journal, but made the mistake of assuming that physicists were
competent to review the paper.  When reputable publications ignore the
existence of computer science, and assume that those without computer
science training are well qualified to do research and review
publications in the area, it hurts the field by allowing outdated or
wrong information to be circulated (in this case, that there is some
sort of breakthrough in what was already a well studied area.)  It is
also insulting.

\bigskip

\noindent {\it Thanks very much to Bob Moore, Chris Meek, Eric Brill, Hagai
Attias, Tom Mitchell, Rich Caruana, Robert Rounthwaite, and Jasha
Droppo, for their discussions on this topic.}

\bibliographystyle{acl}

\bigskip\bigskip
\bigskip
\bigskip

\appendix

\section{Experimental Results}

In this appendix, I briefly describe the experimental results
comparing Naive Bayes to zipping.  The data set used was the 18828
version, available from {\tt
http://www.ai.mit.edu/{\linebreak[0]}\~{}jrennie/{\linebreak[0]}20Newsgroups}.
This version has had duplicates and most headers removed.  No
additional processing of the data was done.  For Naive Bayes, words
were simply segmented at white-space boundaries using the perl
``split'' function.

For Naive Bayes, I computed for each topic the total number of times
each ``word'' occurred in the topic.  I used standard plus-1
smoothing, by, for each topic, adding one count for every word that
occurred in any training set.  The probability assigned to a given
word was thus:
\begin{equation}
\label{eqn:word}
P(\mi{word}|\mi{topic}) = \frac{\mi{count}(\mi{word in topic}) +
 \left\{ \begin{array}{ll}1&\mbox{if \mi{word} occurs in any topic} \\
                         0&\mbox{otherwise} \end{array} \right.}{\mi{total length of topic} + \mi{total number of words that occur in any topic}}
\end{equation}
The count of a word in a topic counts every occurrence of that word in
the topic, counting each word multiple times if it appears multiple
times in the topic.  The total length of a topic counts every
occurrence of every word in the topic, counting multiple times for
words that occur multiple times.  The total number of words that
occur in any topic counts each word that ever appears exactly once, no
matter how many times it appears.

The probability of a document of length $n$ given the topic is simply
$$
P(\mi{document}|\mi{topic}) = \prod_{i=1}^n P(\mi{word}_i|\mi{topic})
$$
If a word occurs more than once in the document, its probability is
multiplied in more than once.  (This estimate neglects the prior on
document length.)

We wish to find the most probable topic, given the document.  The
probability of a topic given a document is found through a simple
application of Bayes' law:
\begin{eqnarray*}
\arg \max_{\mi{topic}} P(\mi{topic}|\mi{document}) & = & \arg \max_{\mi{topic}} \frac{P(\mi{document}|\mi{topic}) \times P(\mi{topic})}{P(\mi{document})} \\
                            & = & \arg \max_{\mi{topic}} P(\mi{document}|\mi{topic}) \times P(\mi{topic})
\end{eqnarray*}
where we can remove the term ${P(\mi{document})}$ because it is
independent of the topic.  Furthermore, we assume a uniform prior on
topics (which is fine for 20 newsgroups, but may be inappropriate for
other tasks.)  Thus, we get
\begin{eqnarray*}
\arg \max_{\mi{topic}} P(\mi{topic}|\mi{document}) & = & \arg \max_{\mi{topic}} P(\mi{document}|\mi{topic})  \\
 & = & \arg \max_{\mi{topic}} -\log P(\mi{document}|\mi{topic}) \\
 & = & \arg \max_{\mi{topic}} -\log \prod_{i=1}^n P(\mi{word}_i|\mi{topic}) \\
 & = & \arg \max_{\mi{topic}} -\sum_{i=1}^n \log P(\mi{word}_i|\mi{topic}) \\
\end{eqnarray*}
Thus, in practice, the algorithm is extremely simple.  For a given
document, for each topic, we simply apply Equation \ref{eqn:word} to
each word in the document; we take the log to avoid underflow; we
compute the sum of those values.  That gives us a value based on the
log of the probability of the topic given the document.  We find the
topic that has the largest negative log probability (which corresponds
to the highest probability.)  Note that all of these techniques are
well known and standard.

For zipping, I applied the algorithm of \namecite{Benedetto:02a}.

The code to split the data into test and training was approximately 70
lines.  The code to implement Naive Bayes was also approximately 70
lines.  No attempt to keep the code small was made.  The code to
implement the zipping algorithm was about 50 lines.  Note that all of
these file sizes are trivial.  For instance, the code I usually use to
do experiments of this nature is over 7000 lines (all written by me.)
It took very roughly an hour to write, debug, and verify the
correctness of each program.  (The zipping program was actually harder
to write, despite its smaller size, because it ran so slowly that it
was more time consuming to debug and verify correctness, and because some
additional effort was made to speed it up, resulting in roughly a
factor of two increase in speed.)

I used the same training and test set for both experiments: I used
99\% of the data for training, and 1\% for testing.  This left 192 test
documents, which is easily large enough to reliably detect errors of
the magnitude seen here.  (I used a relatively small test set because
the zipping algorithm was so slow.)  The Naive Bayes algorithm made 26
mistakes on this test set, a 14\% error rate, while zipping made 91
errors, a 47\% error rate.  The Naive Bayes algorithm required 5
minutes to run, elapsed time, while zipping required one hour and 25
minutes, elapsed time, on the same machine with no other programs in
use.  This was despite minor attempts to optimize the zipping
algorithm and no attempts to optimize the Naive Bayes algorithm.  Note
that the Naive Bayes algorithm was implemented in perl, which is
typically much slower than, say, C or C++.  The zipping script was
also written in perl, but most of the time was spent in gzip, which is
written in C.

Thus, it appears that with approximately equal effort, and using well
known techniques, one can perform text categorization over 3 times
more accurately and 17 times faster by using Naive Bayes rather than
the zipping technique.

\end{document}